\documentclass[aps,prb,twocolumn,groupedaddress]{revtex4-1}
\pdfoutput=1

\usepackage{graphicx}
\usepackage{siunitx}
\usepackage{xcolor}
\usepackage{blindtext}
\usepackage[normalem]{ulem}

\DeclareSIUnit{\elementarycharge}{\text{\ensuremath{e}}}
\DeclareSIUnit{\formulaunit}{\text{\ensuremath{f.u.}}}
\DeclareSIUnit{\bohrmagnetron}{\text{\ensuremath{\mu_\mathrm{B}}}}
\begin{document}

%\preprint{}

\title{Designing magnetism in Fe-based Heusler alloys: a machine learning approach}


\author{Mario {\v Z}ic}
%\affiliation{CRANN and School of Physics, Trinity College Dublin, Dublin 2, Ireland}
\affiliation{School of Physics, AMBER and CRANN Institute, Trinity College, Dublin 2, Ireland}
\author{Thomas Archer}
\email[]{archert@tcd.ie}
\homepage[www.materials-mine.com]{}
%\affiliation{CRANN and School of Physics, Trinity College Dublin, Dublin 2, Ireland}
\affiliation{School of Physics, AMBER and CRANN Institute, Trinity College, Dublin 2, Ireland}
\author{Stefano Sanvito}
\affiliation{School of Physics, AMBER and CRANN Institute, Trinity College, Dublin 2, Ireland}

\date{\today}

\begin{abstract}
Combining material informatics and high-throughput electronic structure calculations offers the possibility of a 
rapid characterization of complex magnetic materials. Here we demonstrate that datasets of electronic
properties calculated at the {\it ab initio} level can be effectively used to identify and understand physical trends 
in magnetic materials, thus opening new avenues for accelerated materials discovery.
Following a data-centric approach, we utilize a database of Heusler alloys calculated at the density functional
theory level to identify the ideal ions neighbouring Fe in the $X_2$Fe$Z$ Heusler prototype.
The hybridization of Fe with the nearest neighbour $X$ ion is found to cause redistribution of the on-site Fe charge 
and a net increase of its magnetic moment proportional to the valence of $X$. Thus, late transition metals are 
ideal Fe neighbours for producing high-moment Fe-based Heusler magnets. At the same time a thermodynamic 
stability analysis is found to restrict $Z$ to main group elements.
Machine learning regressors, trained to predict magnetic moment and volume of Heusler alloys, are used to determine the 
magnetization for all materials belonging to the proposed prototype. We find that Co$_2$Fe$Z$ alloys, and in particular 
Co$_2$FeSi, maximize the magnetization, which reaches values up to \SI{1.2}{T}. This is in good agreement with both 
{\it ab initio} and experimental data. Furthermore, we identify the Cu$_2$Fe$Z$ family to be a cost-effective materials 
class, offering a magnetization of approximately \SI{0.65}{T}.

\end{abstract}

%\keywords{}

\maketitle


\section{Introduction}
Heusler alloys, a vast family of ternary compounds, are often considered an ideal platform for engineering 
and designing novel functional materials. Such class includes both metals and insulators, and among them 
superconductors, topological insulators, thermoelectric alloys, and both optical and magnetic 
materials~\cite{felser_basics_2015, graf_heusler_2011}. As such, the possibility of using alloys of this family for 
fine tuning and controlling the electronic structure and the magnetic order is tantalizing. However, despite 
several decades of intense investigation and accumulated understanding on the Heuslers compounds~\cite{graf_simple_2011}, 
the tuning of their properties still proceeds via chemical intuition in a slow trial-and-error mode. It is then an 
intriguing prospect to explore more high-throughput methods for materials screening and understand whether 
these can identify novel designing rules.

Reliable and low-cost computational methods now allow one to perform systematic investigations of large regions of the 
chemical space. This is known as the computational high-throughput approach~\cite{curtarolo_high-throughput_2013, jain_mgenome_2013}.
The analysis of the generated data has lead data-mining and machine-learning techniques to become part 
of the material science toolbox~\cite{sumpter_perspective_2015, maier_htp_review_2007} 
(for a non-exhaustive list of currently available materials databases see 
[\onlinecite{mmine,materials_project_db, aflowlib_db, harvard_ce_db, oqmd_db, cmrepo_db}]).
Materials can be classified by using 
descriptors~\cite{isayev_mcartography_2015, hachmann_harvard_clean_energy_2011, greeley_ec_for_hydrogen_2006},
simple proxies for sometime complex materials characteristics, and system properties estimated via machine leaning 
regression and 
classification~\cite{shandiz_li_batteries_2016, belisle_ml_interpolation_2015, schutt_rep_crystal_2014, rupp_atomiz_energy_2012}.
The latter are particularly useful when a direct calculation is prohibitive.
An approach based on the machine learning uses statistical inference for predicting properties of a given system without
performing an actual electronic structure calculation. This enables a fast, objective and cost-effective analysis of large amounts 
of multi-dimensional data, making machine learning a natural extension of the computational high-throughput strategy.

\begin{figure}
		\includegraphics[width=1.\columnwidth,
		trim=0 8 0 0, clip]{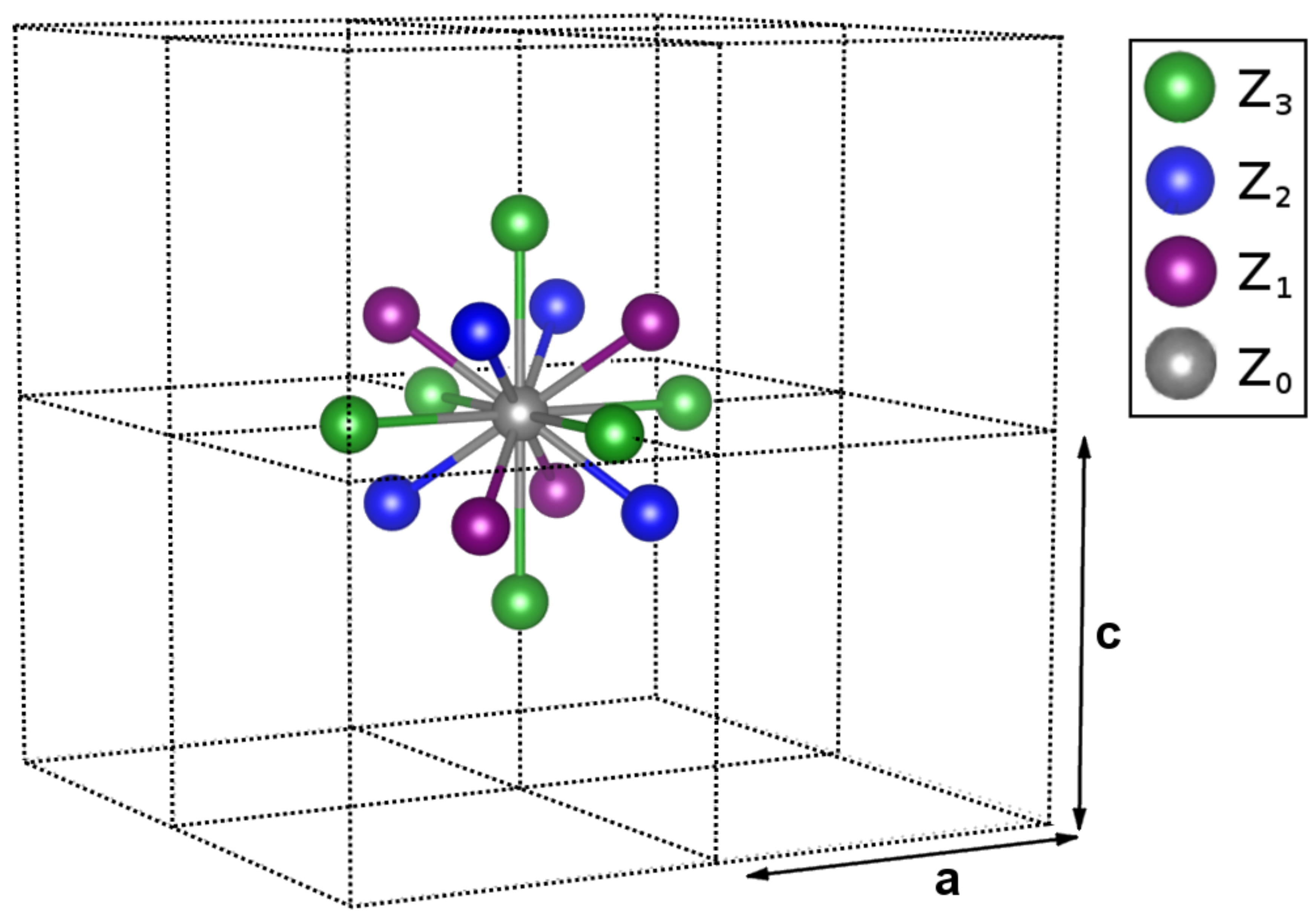}
	\caption{\label{fig_cluster} 
	The local coordination of the atomic sites in a Heulser alloy. The neighbours of the central atom form two shells 
	of different symmetry. Atoms belonging to the nearest neighbour shell, shown in blue and magenta, coordinate the 
	central atom tetrahedrally. The next nearest neighbour shell is made out of six (green) atoms and has octahedral 
	symmetry.
	}
\end{figure}
Here we use machine learning (ML) techniques to study the magnetism of Fe-containing Heusler alloys.
Iron offers a large magnetic moment, second only to Mn, but in contrast to Mn that is known to maintain 
a high-spin state in Heusler compounds~\cite{kubler_Mn_moment_1983}, Fe is more susceptible to changes 
in the local chemical environment. This makes it an ideal choice for exploring the predictive power of ML techniques, 
when applied to magnetic materials. In this work we show that knowing the composition of the first two Fe coordination 
shells is sufficient to accurately estimate its magnetic moment using ML regression. By combining ML and density 
functional theory (DFT) data we are able to identify and explain trends in the compounds magnetic moment. Finally, 
the ideal prototype for an Fe-based regular Heusler alloy is proposed. The ML regression is used to rapidly characterise 
and rank all possible alloys of the proposed prototype, according to the maximal attainable magnetization. We note 
that all predictions are made by only using a list of Fe neighbours, without any need for additional {\it ab initio} calculations.
This demonstrates the potential of the machine learning approach for developing a fast, high-volume, method for 
screening magnetic materials. 

The paper is organized as follows. In the next section we describe the methods at the foundation of the machine-learning
process and the general attributes that enter into the description of magnetism in Heusler alloys. Then we introduce our
results focussing on the role of the nearest neighbour and of next nearest neighbour coordination, and defining the
physical origin of the magnetic moment trends. Then we use our machine learning scheme to identify magnets with large
magnetization. Finally we conclude. 

\section{Method}

\subsection{General Considerations}

Structure-to-property relations are at the heart of all problems in the material science. These are
implicitly determined by the electronic structure of any given compound, which nowadays is routinely 
computed by using {\it ab initio} methods. Accurate information about various material properties can 
then be extracted solely from theory. 
Machine learning (ML) allows us to take a rational approach to large-scale material investigation. The underlying 
assumption is that once there is enough materials data available, an answer to the structure-to-property question 
should be already implicitly contained in the data. We can thus speak about ``learning from the data''.
%However, this approach has not gained a wide acceptance so far.
%
The ML methods are built specifically for this task, providing us with a practical mean to construct approximate 
structure-to-property relationship maps with a well defined domain of validity. The latter, however, needs to be 
established through tests. A major advantage of the ML approach is that it thrives on large datasets, offering a 
high-throughput, objective analysis of the material properties. The data are never discarded, but instead they 
are continuously integrated to refine the predictions and can be reused to address new questions. Thus, it is an 
inductive, data-driven, approach of performing material research. The trade-off is that the ML results are usually 
less accurate than those obtained by using {\it ab initio} methods.

% % describe the main steps of the ML approach
%\subsubsection{general procedure}
%
Here we focus on ``supervised learning'' methods, which include regression and classification algorithms.
The advantage of such class of schemes is that the quality of the ML predictions can be evaluated, for instance 
in the case of the regression by calculating a mean-square error. The purpose of the training procedure is then 
that of minimizing the risk of making incorrect predictions. The idea is that a well-trained algorithm may produce 
a significant error for an individual system but it shall perform in a satisfactory manner for the entire data set.
\begin{figure}
	\includegraphics[width=0.9\columnwidth, height=0.5\columnwidth]{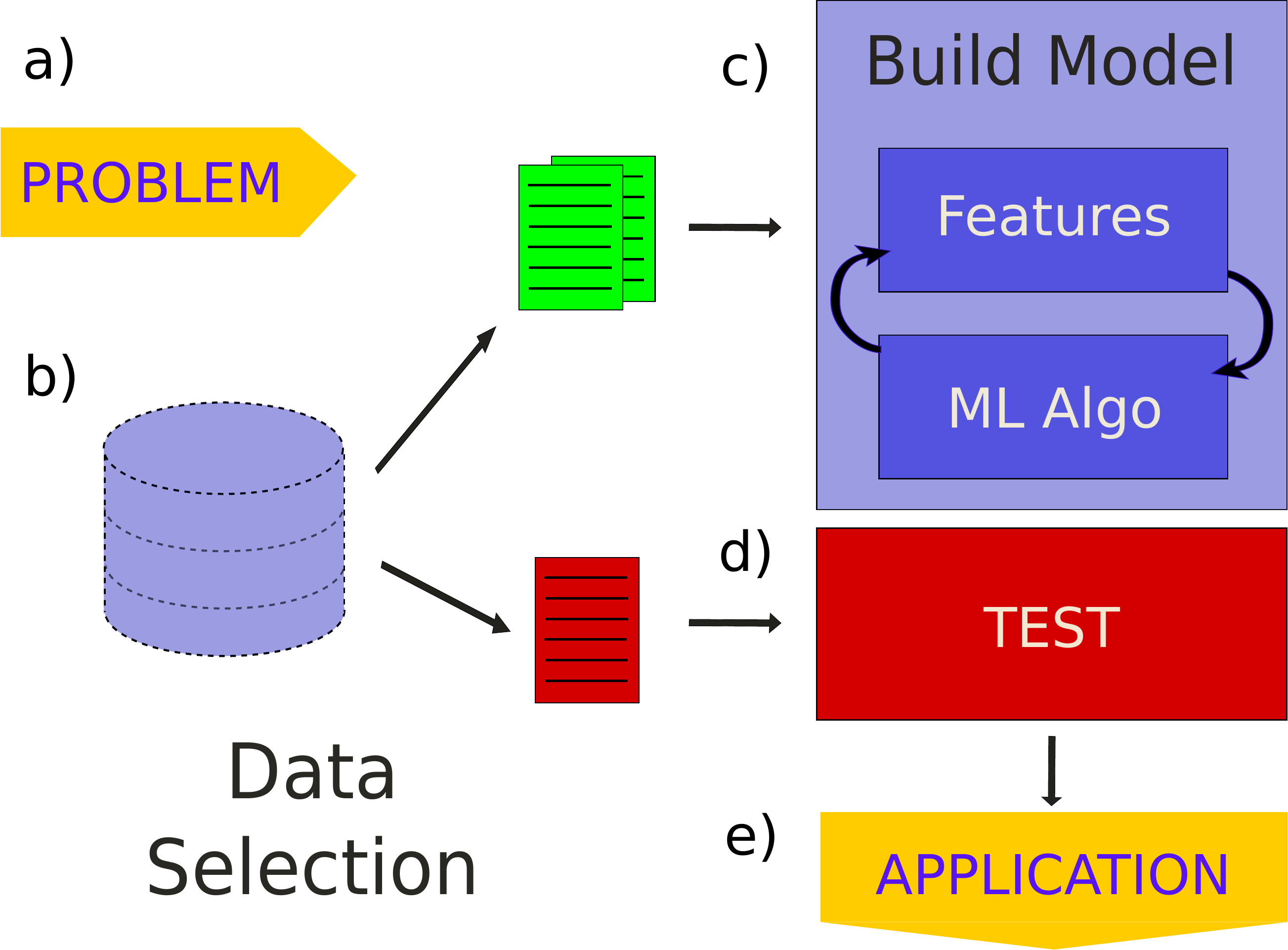}
	\caption{\label{fig_ml_process} 
		Schematic of the supervised learning strategy. The approach consists of 5 steps:
		\textbf{a)} defining the problem, 
		\textbf{b)} selecting and pre-processing the data, 
		\textbf{c)} building the ML model,
		\textbf{d)} testing the model and, finally
		\textbf{e)} applying the ML model to the problem of interest.
	}
\end{figure}
An outline of the general procedure used in this work is shown in figure~\ref{fig_ml_process}. The process starts by 
defining the problem and by selecting suitable data to describe it. The data may come from different sources and is 
usually combined into suitable input features. For example, in figure~\ref{fig_cluster} we use a cluster of atoms to define 
the magnetic moment of the central atom and various atomic data (see~equation~\ref{eq_amag_features}) are 
considered to describe the cluster further.
Next, the available data are split into {the training and the test dataset}, according to the output property that one 
wishes to evaluate. Here it is crucial to preserve the underlying property distribution when the data is split, otherwise one
may end up with a biased {training} set. If the training set is biased, the ML model may not be predictive for data outside
it, and furthermore the fidelity of the algorithm may be erroneously estimated. In brief the ML model trained on such a 
dataset will usually perform badly on new data, i.e. it will not be predictive.
{The test step} is used as an independent 
check of the ML accuracy and its ability to generalize to new data. 
The {test dataset} is never used for building the 
ML model.

The \textit{model building} phase, depicted as a single step in Fig.~\ref{fig_ml_process}, is actually an iterative two-step 
procedure. First a set of input features, namely an input vector encoding a number of chemical/physical properties, is 
constructed from the raw input data and then different ML algorithms are trained using that input. The latter step includes 
the choice of the, e.g., regression algorithm, and the optimization of the hyperparameters. These two steps are repeated 
until a satisfactory accuracy is achieved.
%
%Finally we perform the validation to estimate how the model performs on a new data, which was not used for training.
%

The choice of the input data and its transformation into a useful set of input features is the most important aspect of the 
entire process. In this step we use our domain knowledge to convert the raw data into descriptive features, which correlate 
with the output. This describes an inductive approach for constructing a ML model, which is feasible when we have some 
understanding of the underlying processes correlating the input to the output. In material science this should often be the 
case. A minimal number of input features, which are included in the model, leaves a possibility of its interpretation.
Alternatively, one needs to follow a deductive approach. In this case one starts from a as-large-as possible number of input 
features and performs an input reduction analysis, eliminating the variables that do not correlate with the output. 

There is usually no a unique way to choose the input features and interpret the ML model. If the dimension of the input space 
is not large one may seek to explore the significance of the individual input features. This can help in deepening the knowledge 
of the system under investigation and guide further {\it ab initio} calculations. However, such step is not always possible especially
when the dimension of the input space is large. The importance of having a working ML description of the system is not diminished 
by this feature. It is often much more practical to explore the data using a ML algorithm than working with the raw data. For example, 
in the case of a ML regression, one deals with a single function (a map), instead of a large database. Exploring the connection 
between the variables is thus much simpler and faster. The added benefit is that the ML algorithm will often accurately interpolate 
where data is missing.

\subsection{Machine Learning Model for Magnetism}
In this work we use data extracted from an in-house-made Heusler alloy database, named {\it Materials Mine}~\cite{mmine}.
All calculations have been performed using the PAW~\cite{blochl_paw_1994} pseudopotential implementation of DFT contained
in the VASP code~\cite{paw, vasp, kresse_efficiency_1996, kresse_ab_1994, kresse_ab_1993} and the generalized
gradient approximation of the exchange and correlation functional as parametrized by Perdew, Burke and Ernzerhof~\cite{pbe}.
For each chemical composition, the ground state is calculated for different site occupations, structural parameters and 
sublattice magnetic order. Furthermore, we complement the DFT data by various atomic properties information obtained 
from a wide range of sources in literature, including both experimental and theoretical data. The sources will be properly cited 
individually whenever used.

The knowledge of the crystal cell volume, namely the inter-atomic distance, is vital for studying electronic structure properties
and magnetism in particular. It is reasonable to expect that this quantity will repeatedly appear in all ML models and we wish 
to be able to predict it without relying on the DFT data. We have then trained a volume regressor using the DFT data for the 
{229} fully relaxed full-Heusler structures, having the lowest energy for a given composition and site occupation. We note that 
the number of different compositions in the database is larger than {229}, but we restrict out choice to this dataset since the same
was used with success for other investigations. In any case we will show that this choice does not affect the quality of the final 
result.

The ML model was built using the ridge regressor algorithm as implemented in Scikit-learn package~\cite{scikit-learn}.
We have used \SI{30}{\percent} of the dataset for {the test} and the rest was used as {training set}. 
The input vector was constructed by including the atomic numbers, the atomic volumes, and the atomic radii of the three nonequivalent ions.
The atomic volumes were obtained from Mentel~\cite{mentel2014} and then the atomic radius was calculated for 
each element. Here we assume atoms to be homogeneous solid spheres. A root mean square (RMS) error on {the test 
data} was calculated to be \SI{3.16}{\angstrom\cubed}. In comparison, the mean volume of alloys in the dataset was 
$\approx$~\SI{63}{\angstrom\cubed}. The attained accuracy of the ML algorithm is thus comparable to the precision of the 
DFT calculations, namely is a good predictor for the volume.

We now wish to estimate the magnetic moment of Fe-containing Heusler alloys by using the ML approach. The magnetic 
moment of 3$d$ transition metals is well localized and can be understood as an atomic moment, which gets modified by 
its local surrounding. It is therefore interesting to try to relate the environment of an atom to its moment. The DFT data was 
used to construct clusters of atoms representing the local environment of the central atom, as shown in figure \ref{fig_cluster}.
We construct one cluster for each atomic site of the parent alloy, namely we construct 4 clusters per DFT calculation (per 
Heusler prototype). The data corresponding to the lowest energy states having a given formula unit was selected from the 
database. Note that here we consider a much larger, and less constrained, set of calculations than before. In fact, we 
construct {18,268}{} clusters, of which $\approx$~{7,000}{} were used for {the test}. The site projected magnetic 
moment of the central atom, obtained from the DFT results, was used as the target property.
The input vector was constructed as 
\begin{equation}
\label{eq_amag_features}
\vec{v}_\mathrm{in}=\left(\{Z_i\}, 
	R_0, a_{\mathrm{lat}},
	\{r_{0i}\},
	\{N_i\}, 
	S_0
\right)\:,
\end{equation}
where $\{Z_i\}$ ($i=0,1,2,3$) and $R_i$ are the atomic number and the atomic radius of the $i$-th atom, respectively, and 
$N_i$ is the valence ($i=0,1,2$). The atomic positions are labeled as in Fig.~\ref{fig_cluster}. Here $r_{0i}$ is the 
distance between 0-th and the $i$-th atom, scaled by the sum of their atomic radii. The ``effective'' cubic lattice constant, 
$ a_{\mathrm{lat}}$, is calculated from the volume of the parent Heusler structure, which in turn is estimated using the previously 
discussed regression model. Finally, $S_0$ is the Stoner parameter of the central atom, obtained from 
Janak~\cite{janak_1977}.

As for as the regression is concerned, we have found the Random Forest Regression, as implemented in the 
Scikit-learn library~\cite{scikit-learn}, to give the best results. The RMS error measured on {the test dataset} 
is \SI{0.4}{\bohrmagnetron}, with the RMS error on the subset having Fe as central atom being measured somewhat 
higher, $\sim$~\SI{0.54}{\bohrmagnetron}. We did not notice any improvement in the RMS error when we performed 
the training by only using Fe-centered clusters. We believe there are two reasons for this finding. The first is the reduced 
dataset size utilized. We find that $\sim$~\SI{7 000}{} clusters are barely sufficient to converge the algorithm learning curve.
The second is the that the larger dataset also contains non-magnetic atoms, whose magnetic moment is trivial to estimate.
We note that in some cases the errors are much larger than the RMS value, which seems to be at least partly related to the 
observed convergence issues in the high-throughput calculations. For example, the DFT data is found to exhibit a large variation 
in the magnetic moment of Fe (see Fig.~\ref{fig_effect_of_z3}). Reliable methods for data cleaning are needed, however, in this 
work we treat such anomalous calculations simply as a noise.

\section{Results}

\subsection{Role of the Coordination Shells}
In the Methods section we have presented a ML model for predicting the magnetic moment of an atom embedded in a 
Fe-containing regular Heusler alloy (see figure~\ref{fig_cluster}). The model estimates the magnetic moment based on 
four key variables: the three atomic numbers specifying the coordination of the central atom, $\{Z_i\}$, and the lattice 
constant of the parent Heusler alloy. The latter can be estimated by using a ML regression (see~Table~\ref{tbl_estimates}).
The atomic properties, needed to construct the input vector (Eq.~\ref{eq_amag_features}), are easily obtained.
This makes the method completely free of input {\it ab initio} parameters.
We then use this model to explore and gain a deeper understanding of our DFT data. Here we focus on $X_2$Fe$Z$ alloys, 
where the central atom is Fe, with the cubic L2$_1$ crystal symmetry and the corresponding tetragonal 
structures~\cite{graf_crystal_2009}. In this case the two inequivalent sites in the first coordination shell are occupied by identical 
atoms (i.e. ${Z}_1 ={Z}_2 = {X}$ in Fig.~\ref{fig_cluster}), a fact that allows us to explore the effects of three structural parameters.

\subsubsection{Role of the Next-Nearest Neighbour}
\begin{figure}
		\includegraphics[width=0.85\columnwidth]{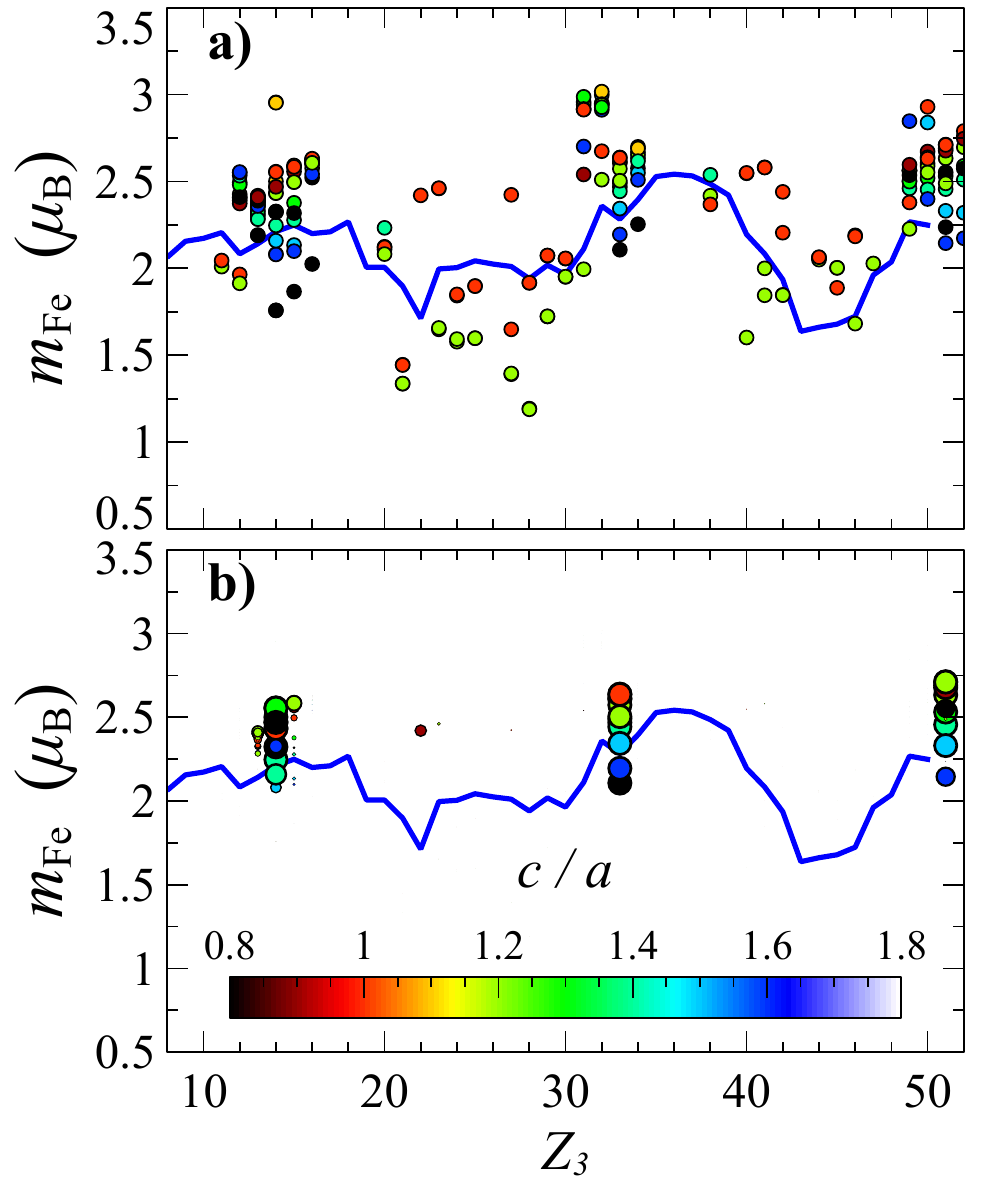}
	\caption{\label{fig_effect_of_z3}
	Estimate of the magnetic moment of Fe, $m_\mathrm{Fe}$. a) $m_\mathrm{Fe}$ (in $\mu_\mathrm{B}$) as a 
	function of the next-nearest-neighbour atomic number, $Z_3$, for a Wigner-Seitz radius of \SI{2.7}{\bohr}.
	The $c/a$ ratio of the parent Heusler alloy is color coded. A machine learning estimate is shown by a solid blue line.
	b) The same data as in a), where now the size of the symbols~\protect\footnote{See Ref.~[\onlinecite{enthalpy_radius}] 
	for details.} is proportional to the calculated enthalpy of formation, $\Delta H$.	 Large circles correspond to more stable 
	alloys.
%	\textcolor{red}{TODO: make the figures bigger by putting them on top of each other. \textcolor{blue}{STEFANO: use $m_\mathrm{Fe}
%	(\mu_\mathrm{B})$ as label for the $y$ axis, and use $Z_3$ (italic) for the $x$ one. Magnify all symbols, these are really small.}}
	}
\end{figure}	
We first look at the composition of the second coordination shell, i.e. the effect of $Z_3$ of the magnetic moment of Fe,
$m_\mathrm{Fe}$ (values are provided in unit of Bohr magneton, $\mu_\mathrm{B}$). The DFT dataset was sampled at 
a constant volume, namely the Wigner-Seitz radius was set to $R_\mathrm{WS}=2.7 \mathrm{\si{\bohr}}$, and the 8 nearest 
neighbour atoms were fixed to Fe. This Wigner-Seitz radius roughly corresponds to a Heusler lattice constant of \SI{5.8}{\angstrom}. 
The corresponding data and the ML estimate of the magnetic moment are shown in Fig.~\ref{fig_effect_of_z3}(a).

Visual inspection of the data reveals that transition metals ($21<Z_3<30$, $39<Z_3<48$) and main group elements 
($13<Z_3<16$, $31<Z_3<34$, $49<Z_3<52$) make two distinct classes of next-nearest-neighbours. On the one hand,
transition metals tend to increase the magnetic moment of Fe proportionally to their valence. On the other hand, main 
group elements tend to cluster and yield a maximal magnetic moment. In particular, for a given nearest 
neighbour and volume $m_\mathrm{Fe}$ is only weakly affected by the choice of the main group element at the 
$Z_3$ site. This can be seen, for example, in figure~\ref{fig_fe_tm_all_volume_dep}, where the magnetic moment
of Fe appears not to be correlated to the atomic number of the next-nearest neighbour, $Z_3$.
Here we would like to point out that this does not contradict the 
well established Slater-Pauling rule~\cite{galanakis2002, graf_simple_2011}.
In particular, for Co$_2XY$ alloys this rule would imply that the net cell moment, $m_\mathrm{cell}$,
scales linearly with the valence of the alloy, $N_\mathrm{V}$, and reaches maximum when $N_\mathrm{V}=30$.
As an example, for Co$_2$FeAl, Co$_2$FeSi and Co$_2$FeP alloys ($N_\mathrm{V}=$~29, 30 and 31)
we find the DFT moments, $m_\mathrm{cell}$, to be \SIlist{5.10;5.48;4.68}{\bohrmagnetron} (per \si{\formulaunit}), respectively.
However, the corresponding Fe moments change only slightly,
namely we find $m_\mathrm{Fe}=$~\SIlist{2.77;2.80;2.65}{\bohrmagnetron}.

We note that the observed ML trend is volume dependent, giving a spurious representation of the valence trend across 
the transition metal series. The problem, however, does not affect the main group elements. The ML magnetization trend 
captured at smaller volumes, $R_\mathrm{WS} \approx 2.4 \mathrm{\si{\bohr}}$, is qualitatively different from the trend 
at larger ones, $R_\mathrm{WS} \approx 2.8 \mathrm{\si{\bohr}}$. In contrast, the DFT data trends shown in Fig.~\ref{fig_effect_of_z3} 
remain by large unaffected by the volume change. We, therefore, find it necessary to combine the DFT data and the ML 
approach to obtain a complete picture. In spite of this, the numerical precision of the ML estimate is always within the limits 
established by the regressor test procedure, $\sim \SI{0.5}{\bohrmagnetron}$.

From a material design perspective, the learning is that the $Z_3$ element can be chosen to ensure the stability of a 
Heusler alloy without compromising the magnetic moment. Figure \ref{fig_effect_of_z3}(b) shows the enthalpy of formation 
for the same set of Heusler alloys shown in figure \ref{fig_effect_of_z3}(a). The enthalpy of the alloy is calculated with 
respect to the decomposition into the most stable elemental phases, so that does not provide a strict stability criterion, 
but simply a guideline for stability~\cite{ourpaper}. With a small number of exceptions we find that only the main group 
elements at the $Z_3$ site have a good chance to yield thermodynamically stable alloys. Such a result could be anticipated 
based on the known Heusler chemistry~\cite{graf_simple_2011}. The freedom of the choice of $Z_3$ opens up the possibility 
to tune the volume of the alloy, and to control the critical temperature~\cite{shirakawa_tc_pressure_1987, adachi_tc_Rh2MnZ_2004}.
In Table~\ref{tbl_estimates} we show that Heusler alloy volume can be accurately estimated using the ML regression.

\subsubsection{Role of the Nearest Neighbour}
\begin{figure}
		\includegraphics[width=1.0\columnwidth]{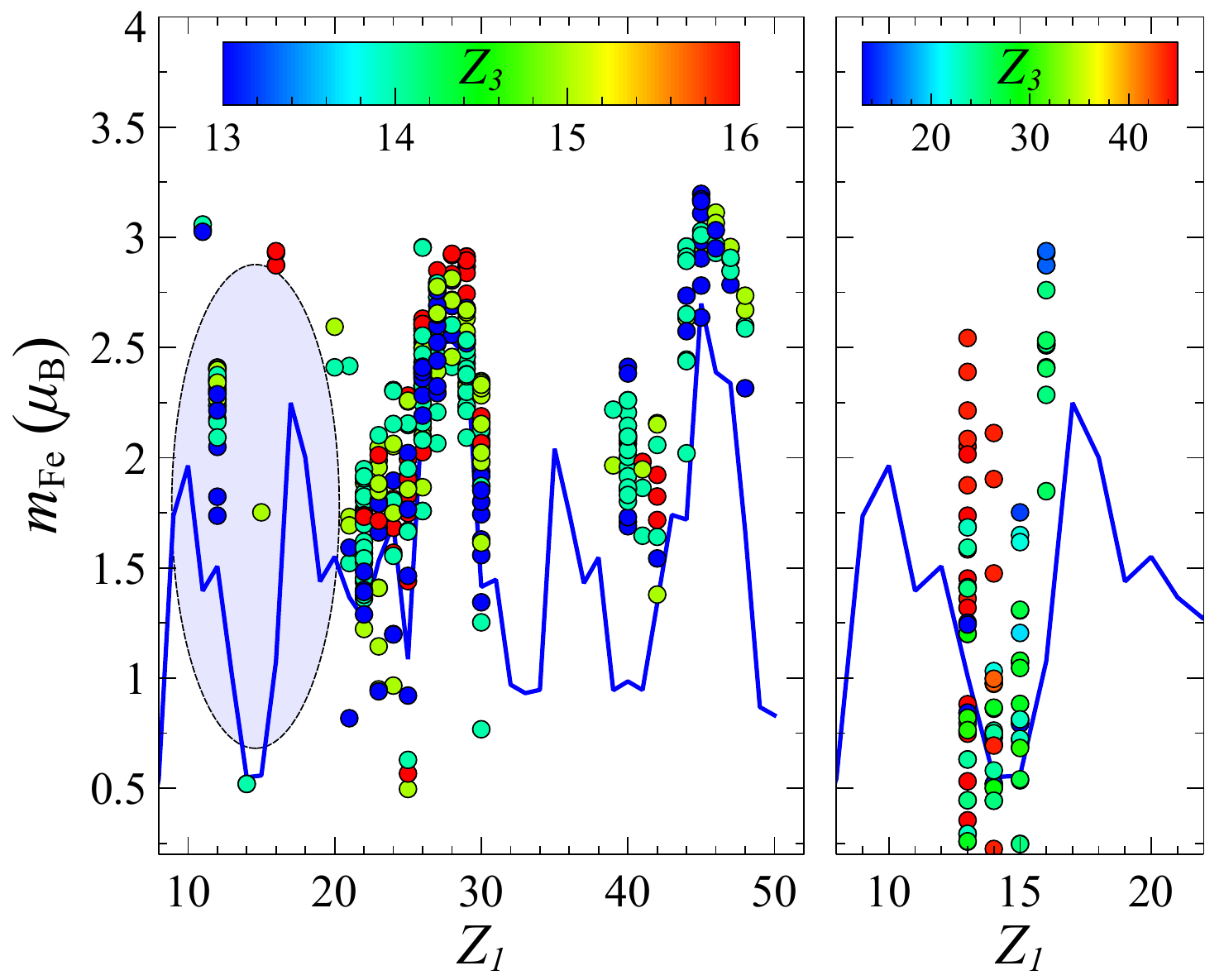}
	\caption{\label{fig_fe_tm_all_volume_dep} 
	\textit{Left panel} - Magnetic moment moment of Fe, $m_\mathrm{Fe}$ (in $\mu_\mathrm{B}$), for a wide range of 
	nearest neighbours at a constant Wigner-Seitz volume {($R_\mathrm{WS} =$~\SI{2.7}{\bohr})}. 
	The atomic number of the next nearest neighbour, $Z_3$, is color coded, while here we plot data as a function of
	the first nearest neighbour atomic number, $Z_1$. We can notice a linear increase of the magnetic moment across the 
	transition metal series which does not depend on $Z_3$. The symbol are the DFT data while the corresponding machine 
	learning trend is shown with {the blue line}. 
	\textit{Right panel} - a data sample containing a wider range of main group elements.
	The data elucidates the origin of the oscillation in the machine learning trend throughout the main group series.
%	\textcolor{blue}{STEFANO: use $m_\mathrm{Fe}$ ($\mu_\mathrm{B}$) for the $y$ axis, $Z_1$ for the $x$ and 
%	$Z_3$ for the color thermometer. The inset is really difficult to read, maybe you can take it out and plot this as a 
%	two-panel picture. In general the quality of the figure is not good.}
	}
\end{figure}
%%
%%
%\textcolor{red}{EDIT: }
In the previous discussion we have shown that the magnetic moment of Fe is independent of the choice of $Z_3$ as long 
as $Z_3$ is a main group element. We now study the magnetic moment as a function of the nearest neighbour ion, $Z_1$,
keeping a main group element at the $Z_3$ site and a fixed volume (see Fig.~\ref{fig_fe_tm_all_volume_dep}). The trend 
with volume of $m_\mathrm{Fe}$ depends on the choice of Z$_1$ and it is difficult to qualify. The ML regression, however, 
can be used to take the volume effects into account with a good level of precision. We find that the valence of the nearest 
neighbour ions determines the magnetic moment of Fe for the entire range of volumes investigated. When the valence of 
the nearest neighbour ion is less than 8 the moment decreases, and conversely, when it is larger it increases. The same 
trend is found for all nearest neighbours belonging to the 3$d$ and the 4$d$ transition metal series, as shown in 
Fig.~\ref{fig_fe_tm_all_volume_dep}. The maximal moment of Fe is obtained when Ni or Pd constitute the first coordination 
shell. The ML trend, shown with {a blue line}, reproduces the magnetization trend for all possible nearest neighbours.
For main group neighbours the magnetization shows a strong variation with the valence, taking a minimum value in the middle 
of the series. The overall magnetic moment is then reduced when compared to the situation with transition metal elements . 
In conclusion, we have found that late transition metals: Co, Ni, Cu, Rh, Pd, and Ag, make the most desirable nearest neighbours 
of Fe, since they maximize its local magnetic moment.

\subsubsection{Physical Origins of the Magnetic Moment Trends}
\begin{figure}
		\includegraphics[width=1.0\columnwidth]{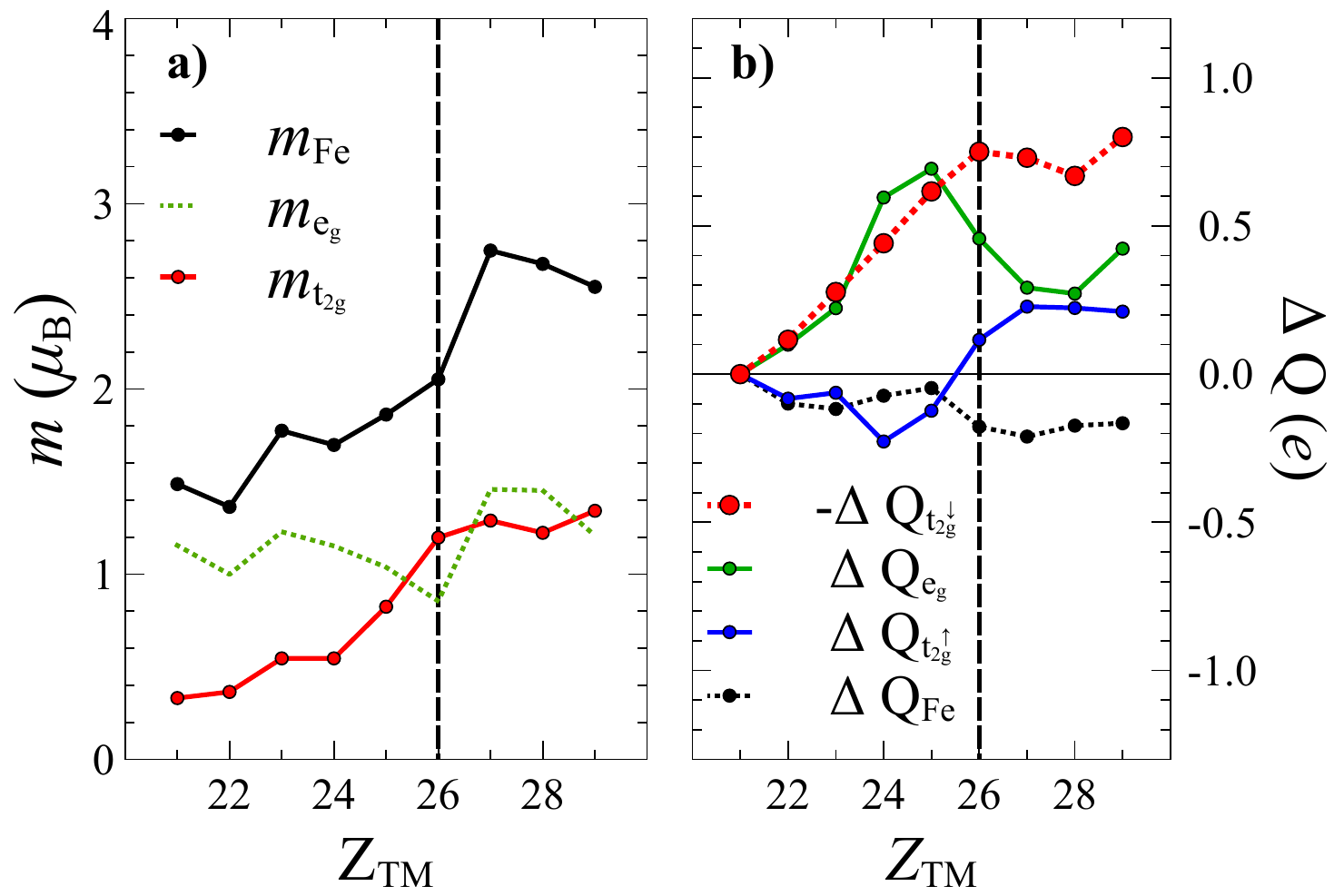}
	\caption{\label{fig_fe_z1_charge} 
	\textbf{a)} Orbital resolved magnetic moment of Fe, $m_\mathrm{Fe}$, for TM$_2$FeZ clusters,
	where the selected TMs are 3$d$ elements ($Z_\mathrm{TM} = 21$ to $Z_\mathrm{TM} = 29$).
	The magnetic moment of the e$_\mathrm{g}$ band, $m_{\mathrm{e}_\mathrm{g}}$, 
	is roughly constant throughout the series, while the total moment, $m_\mathrm{Fe}$, 
	increases following the increase of the t$_{2\mathrm{g}}$ moment, $m_{\mathrm{t}_\mathrm{2g}}$.
	\textbf{b)} Orbital and spin resolved change of the Fe site projected charge, $\Delta \mathrm{Q}_\mathrm{Fe}$,
	throughout the TM series.
	As a reference we take the site projected charges of Sc$_2$FeSi ($Z_\mathrm{TM} = 21$).
	For the t$_{2\mathrm{g}}$ minority spin band we plot the charge loss, $- \Delta \mathrm{Q}_\mathrm{t_{2g}^{~\downarrow}}$,
	to clearly show its correlation with the charging of the Fe $e_g$ band, $\Delta \mathrm{Q}_\mathrm{e_{g}}$.
	}
\end{figure}
In order to understand the physical origin of the established trend in the transition metal series, we select a subset of the compounds 
shown in Fig.~\ref{fig_fe_tm_all_volume_dep} for further analysis. In particular we look at the following Fe-containing alloys: 
Sc$_2$FeSi, Ti$_2$FeAl, V$_2$FeSi, Cr$_2$FeSi, Mn$_2$FeS, Fe$_2$FeAl, Co$_2$FeAl, Ni$_2$FeSi and Cu$_2$FeSi, which 
all possess a cubic L2$_1$ structure. We note that, in general, the tetragonal distortion does not change the main trend, so that
it is not considered here. An analysis of the site projected density of states (PDOS) reveals that both the Fe and the nearest neighbor 
atom remain charge neutral throughout the series. This means that Fe is always occupied by $6$ electrons and the origin of the 
magnetic moment trend is solely due to the on-site charge redistribution, as shown in figure~\ref{fig_fe_z1_charge}.
By integrating the orbital resolved Fe PDOS we find that for early TMs ($Z_\mathrm{TM} \leq 25$)
the amount of charge transferred from the minority $t_{2\mathrm{g}}$ band to the $e_\mathrm{g}$ spin bands is proportional to the 
nearest neighbour valence, with the charge of the majority $t_{2\mathrm{g}}$ band, located deep below the Fermi level, remaining 
roughly the same. As a result, the Fe net magnetic moment increases proportionally to the valence of the nearest neighbour.
For Mn, Fe and Co nearest neighbour we observe a reverse charge transfer, from the minority $e_\mathrm{g}$ band to the majority 
$t_{2\mathrm{g}}$ one, $\Delta \mathrm{Q_{t_{2g}^{~\uparrow}}} \approx$~\SI{0.4}{\elementarycharge}, resulting in an noticeable 
kink in the magnetic moment trend.

\begin{figure}
		\includegraphics[width=1.0\columnwidth]{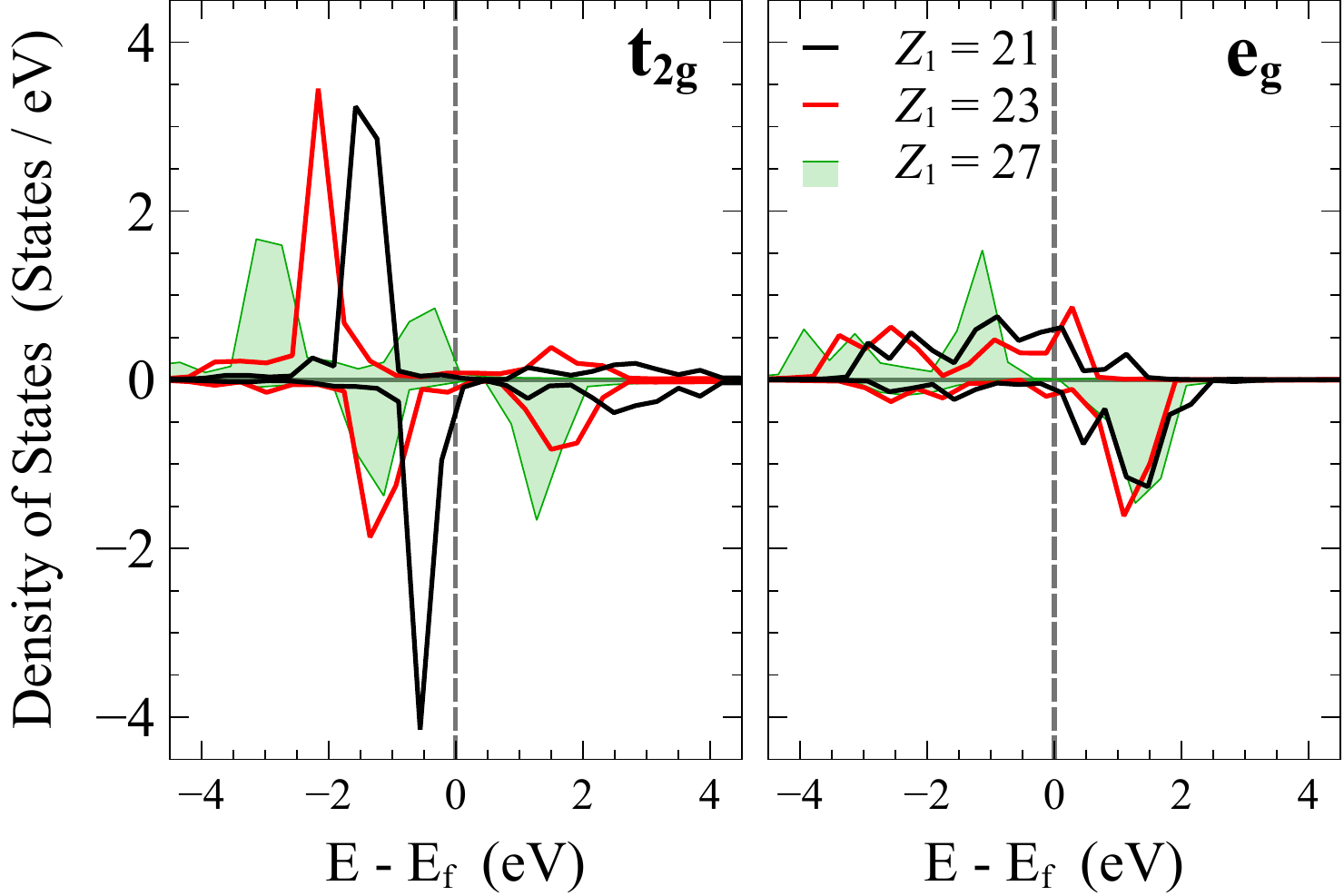}
	\caption{\label{fig_fe_z1_pdos} 
	Projected density of states of Fe in the TM$_2$FeZ clusters, where the selected TMs are Sc ($Z_1=21$), V ($Z_1=23$) 
	and Co ($Z_1=27$). The t$_{2\mathrm{g}}$ band is shown in the left-hand side panel and the e$_{\mathrm{g}}$ in the 
	right-hand side one. The spin up and down channels are shown as the positive and the negative values, respectively.}
\end{figure}
Our understanding of such charge re-distribution mechanism is the following. At the begining of the transition metal series, the 
nearest neighbour atoms hybridize weakly with the $t_{2\mathrm{g}}$ band of Fe, resulting in a narrow and strongly spin split 
Fe $t_{2\mathrm{g}}$ band {(see Fig.~\ref{fig_fe_z1_pdos})}. 
As such the energy overlap of the Fe $t_{2\mathrm{g}}$ spin bands is initially very small. By increasing 
the valence of the nearest neighbour ion one leads to a stronger hybridization with the Fe $t_{2\mathrm{g}}$ spin bands, which
then get wider. Consequently, their overlap increases, giving rise to an increasingly strong repulsive Coulomb interaction. It is 
therefore energetically more favourable to transfer a part of the minority $t_{2\mathrm{g}}$ charge to the $e_\mathrm{g}$ band,
which is strongly spin-split and \SIrange{5}{6}{\electronvolt} wide in energy, indicating spatially delocalized orbitals.
For late TM neighbours, e.g.~Co ($Z_1 = 27$), the hybridization with Fe results in a wide majority $d$-band,
which can now accommodate extra electrons since the Coulomb repulsion is reduced by the band broadening.
At the same time the energy cost of adding extra electrons to the $e_{\mathrm{g}}$ band is increased, as the band 
is nearly full. For Mn neighbours ($Z_1 = 25$) we find the Fe $e_\mathrm{g}$ charge to be 
Q$_\mathrm{e_g}~\approx$~\SI{2.5}{\elementarycharge}. The reverse charge transfer thus reduces the Coulomb 
energy of the $e_\mathrm{g}$ band. The $d$-band electronic structure of Fe in Sc$_2$FeSi, V$_2$FeSi and 
Co$_2$FeSi alloys, which clearly illustrate the mechanism just described, is shown in Fig.~\ref{fig_fe_z1_pdos}.

\subsection{Application}
\subsubsection{Screening of High-Magnetic-Moment Heulser alloys using ML Methods}
The analysis carried out on the dependence of the Fe magnetic moment on the local chemical environment enables 
us to propose \textit{LTM}$_2$Fe\textit{MG} as an optimal chemical composition for a ternary Fe-based Heusler alloy
with maximum $m_\mathrm{Fe}$ ({\it LTM} stands for late transition metal and {\it MG} for main group element).
We stress that our assertion applies only to  L$2_1$-type Heusler alloys, but notably a well-established preferential 
site occupation rule~\cite{graf_crystal_2009}, suggests that the proposed stoichiometry will crystallize in the required 
regular Heusler phase.
\begin{table}
\caption{\label{tbl_estimates} The volume and the magnetic moment of Fe for a number of Heusler alloys. The machine 
learning (ML) estimates are compared to the DFT results. Volumes are given for the primitive unit cell containing 4 atoms.
The length (volume) is expressed in \si{\angstrom} ($\mathrm{\si{\angstrom}}^3$) and the magnetic moments are in 
\si{\bohrmagnetron}.}
\vspace{5pt}
\begin{ruledtabular}
\begin{tabular}{ccccccc}
Compound & $a_{||}$ & $c/a$ & V$_{\mathrm{DFT}}$ & V$_{\mathrm{ML}}$ & M$_{\mathrm{DFT}}$ & M$_{\mathrm{ML}}$ \vspace{3pt}\\ 
Co$_2$FeSi & 5.63 & 1.2 & 44.49 & 45.86 & 2.79 & 2.69 \\
Cu$_2$FeAl & 5.51 & 1.2 & 50.20 & 45.44  & 2.52& 2.67 \\
Rh$_2$FeSn & 5.89 & 1.2 & 61.21 & 59.74 & 3.13 & 3.12 \\
Ni$_2$FeAl & 5.39 & 1.0 & 47.02 & 43.77 & 2.69 & 2.69 \\
Ni$_2$FeGa & 5.38 & 1.2 & 46.70 & 47.31 & 2.73 & 2.81 \\
\end{tabular}
\end{ruledtabular}
\end{table}
There exists a number of Heusler alloys reported in literature, which belong to the proposed prototype. For example, Co$_2$FeSi is 
a well known ferromagnetic half-metal with a critical temperature of \SI{1100}{\kelvin}~\cite{wurmehl_Co2FeSi_2006}. Other examples 
of related Heusler alloys include: Co$_2$FeAl, Cu$_2$FeAl, Ni$_2$FeAl, Ni$_2$FeGa and 
Rh$_2$FeSn~\cite{kobayashi_Co2FeAl_2004, zhang_Cu2FeAl_2004, zhang_Ni2FeAl_2007, liu_Ni2FeGa_2003, suits_heusler_1976},
proving that the proposed prototype has a good chance to yield thermodynamically stable alloys.

The structural and the magnetic properties of these alloys can be predicted by using the machine learning regression.
In Table~\ref{tbl_estimates} we compare the ML results for the volume and the Fe magnetic moment for the aforementioned 
alloys with the corresponding DFT values extracted from the DFT database~\cite{mmine}, demonstrating indeed a good agreement.
We note that the method presented here can only be used to directly evaluate the magnetization of alloys containing a single Fe 
atom and no other magnetic elements. When other magnetic ions are present in the composition the ML method will in general 
underestimate the total cell magnetic moment, and it will need to be extended to take the magnetic ordering into account.
The fact that most of these Heusler type present ferromagnetic ordering (a ferromagnetic ground state is found to be stable over the 
entire range of volumes, namely $R_\mathrm{WS}=\mathrm{\SI{2.3}{\bohr}~to~\SI{3.0}{\bohr}}$) allows us to easily account for the 
additional magnetic atoms. In fact the total moment per cell can be obtained by simply adding the average magnetic moment of the 
LTMs to the magnetic moment of Fe.

\begin{figure}
		\includegraphics[width=0.9\columnwidth]{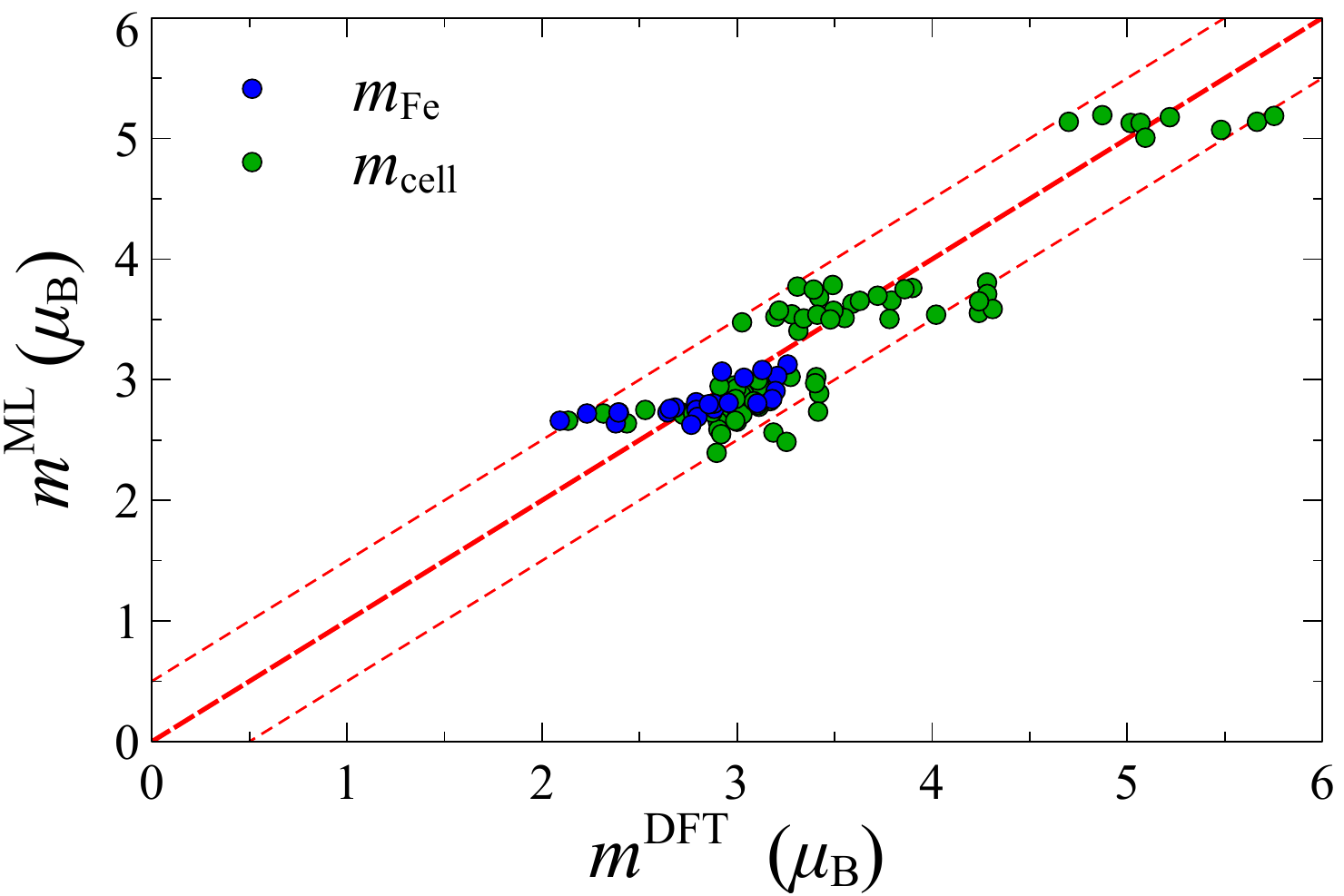}
	\caption{\label{fig_ltm2femg_estimates}
	Comparison between the magnetic moments of \textit{LTM}$_2$Fe\textit{MG} Heusler alloys predicted by the ML 
	regression, $m^\mathrm{ML}$, and those calculated with DFT, $m^\mathrm{DFT}$. The magnetic moment of Fe, 
	$m_\mathrm{Fe}$, is predicted directly using the regression (blue dots). The total magnetic moment per cell, 
	$m_\mathrm{cell}$, (green dots) is estimated by using the empirical correction scheme described in the text. 
	The red line denotes perfect agreement, $m^\mathrm{ML}=m^\mathrm{DFT}$.
	}
\end{figure}
Here among the LTMs only Co, Ni, Rh, and Ir are found to have appreciable magnetic moments and their average values, $\bar{m}$, 
have then been estimated by using the site-projected magnetic moment data of various Heusler alloys found in the database. We estimate 
the following average moments: $\bar{m}_\mathrm{Co}=1.19~\mu_\mathrm{B}$, $\bar{m}_\mathrm{Ni}=0.37~\mu_\mathrm{B}$,
$\bar{m}_\mathrm{Rh}=0.34~\mu_\mathrm{B}$ and $\bar{m}_\mathrm{Ir}=0.35~\mu_\mathrm{B}$. The remaining late transition 
metals tend to be either non-magnetic or weakly magnetic, leaving Fe as the only source of magnetic moment.

We have then used the method described to characterize all the possible compounds of the proposed \textit{LTM}$_2$Fe\textit{MG} 
prototype. The main group elements have been chosen among: Al, Si, P, Ga, Ge, As, In, Sn and Sb, and the late transition metals 
among: Co, Ni, Cu, Rh, Pd, Ag, Ir, Pt and Au. The magnetization has been calculated for each Heusler alloy by using the ML estimate 
of the magnetic moment and the volume. The results have been compared \textit{a posteriori} with the DFT ones and found to be in 
a good agreement, see Fig.~\ref{fig_ltm2femg_estimates}. The error for the magnetic moment is below 
\SI{0.5}{\bohrmagnetron \per \formulaunit} for all the alloys considered. We have also found that Rh and Ir are the two ions, which allow
one to maximize the Fe magnetic moment, reaching out a value of \SI{3}{\bohrmagnetron \per atom}. However, the maximal 
cell magnetization of \SI{1.2}{\tesla} was achieved in Co$_2$-based magnets, with Ni$_2$- and the Cu$_2$-based based alloys 
following and having a magnetization of \SI{0.83}{\tesla} and \SI{0.65}{\tesla}, respectively.

\section{Conclusions}
In conclusion we have investigated the magnetic moment of Fe in Heusler alloys and its dependence on the local chemical 
environment. We have identified the valence of the Fe neighbours as the key parameter governing the moment. The 
\textit{LTM}$_2$Fe\textit{MG} prototype has been found to be the ideal for an Fe-based ternary Heusler alloy with maximum
magnetization. By using machine learning algorithms we have estimated the volume and the magnetic moment for the entire 
family of such compounds, and the alloys have been ranked according to their performance, namely the maximal magnetization.
We find Co$_2$FeSi and Co$_2$FeAl at the top of our list. These are a well known high-performance magnets for 
spintronics~\cite{wurmehl_Co2FeSi_2006, kobayashi_Co2FeAl_2004}. For large-scale production or in applications as
permanent magnets, where the performance is measured in magnetization per dollar, Cu$_2$-based magnets, such as 
Cu$_2$FeAl~\cite{zhang_Cu2FeAl_2004}, become the best choice.
Finally, we have demonstrated that machine learning can be used as a cost-effective and reliable method for material 
characterization. We have also shown that combining material informatics and high-throughput DFT calculations makes 
a powerful platform for accelerated materials research.

\begin{acknowledgments}
This work has been funded by Science Foundation Ireland (Grant No. 14/IA/2624 and AMBER Center).
Computational resources have been provided by the Trinity Center for High Performance Computing (TCHPC) 
and by the Irish Centre for High-End Computing (ICHEC).
\end{acknowledgments}

\bibliography{Manuscript.bib}

\begin{thebibliography}{43}
\bibitem [{ent()}]{enthalpy_radius}%
  \BibitemOpen
  \href@noop {} {}\bibinfo {note} {{The radius of the point is calculated as
  $r=\exp{\left(-\frac{\Delta H}{\mathrm{\SI{0.2}{\electronvolt}}}\right)}$,
  where $\Delta H$ is the enthalpy of formation.}}\BibitemShut {Stop}%
\end{thebibliography}

\end{document}